\def\gsim{\;\lower4pt\hbox{${\buildrel\displaystyle >\over\sim}$}\,}
\def\lsim{\;\lower4pt\hbox{${\buildrel\displaystyle <\over\sim}$}\,}
\def\em{{\rm em}}
\def\chandra{{\it Chandra}}
\def\swift{{\it Swift}}
\def\FLASH{{\sc flash}}
\def\PARAMESH{{\sc paramesh}}

\documentclass{aa}  
\usepackage{graphicx}
\usepackage{natbib}
\usepackage[english]{babel}
\usepackage{txfonts}
\newcommand\rs[1]{_\mathrm{#1}}

\begin{document}
   \title{Three-dimensional modeling of the asymmetric blast wave from
          the 2006 outburst of RS Ophiuchi: Early X-ray emission}

   \author{S. Orlando\inst{1,2},
           J.J. Drake\inst{3},
          \and
           J.M. Laming\inst{4}
          }

   \offprints{S. Orlando,\\ e-mail: orlando@astropa.inaf.it}

   \institute{INAF - Osservatorio Astronomico di Palermo ``G.S.
              Vaiana'', Piazza del Parlamento 1, 90134 Palermo, Italy
         \and
              Consorzio COMETA, via Santa Sofia 64, 95123 Catania, Italy
         \and
              Harvard-Smithsonian Center for Astrophysics, 60 Garden
              Street, Cambridge, MA 02138, USA
         \and Space Science Division, Naval Research Laboratory,
              Code 7674L, Washington DC 20375, USA
             }

   \date{Received \quad\quad\quad ; accepted \quad\quad\quad }

   \authorrunning{S. Orlando et al.}
   \titlerunning{3D modeling of the asymmetric blast wave from RS Ophiuchi}

 
  \abstract
   {\chandra/HETG observations of the recurrent nova RS Ophiuchi at day
    13.9 of its 2006 outburst reveal a spectrum covering a large range
    in plasma temperature and characterized by asymmetric and blue-shifted
    emission lines (\citealt{2008ApJ...673.1067N, drake07}).}
   {We investigate the origin of asymmetries and broadening of the
    emission lines observed with \chandra/HETG. We explore possible
    diagnostics of the early blast wave and of the circumstellar medium
    (CSM) in which the explosion occurred.}
   {We perform 3-D hydrodynamic simulations of the blast wave from the
    2006 outburst, propagating through the inhomogeneous CSM. The model
    takes into account the thermal conduction (including the effects of
    heat flux saturation) and the radiative cooling. From the simulations,
    we synthesize the X-ray emission and derive the spectra as they
    would be observed with \chandra/HETG.}
   {The simulated nova remnant is highly aspherical and the blast wave is
    efficiently collimated by the inhomogeneous CSM. Our model reproduces
    the observed X-ray emission in a natural way if the CSM in which
    the outburst occurred is characterized by an equatorial density
    enhancement. Our ``best-fit'' model predicts that most of the
    early X-ray emission originates from a small region propagating in
    the direction perpendicular to the line-of-sight and localized just
    behind the interaction front between the blast wave and the equatorial
    density enhancement. The model predicts asymmetric and blue-shifted
    line profiles remarkably similar to those observed. These asymmetries
    are due to substantial X-ray absorption of red-shifted emission by
    ejecta material.}
   {The comparison of high quality data of \chandra/HETG with detailed
    hydrodynamic modeling has allowed us to unveil, for the first
    time, the details of the structure emitting in the X-ray band in
    early phases of the outburst evolution, contributing to a better
    understanding of the physics of interactions between nova blasts
    and CSM in recurrent novae. This may have implications for
    whether or not RS Ophiuchi is a Type Ia SN progenitor system.}

   \keywords{shock waves --
             binaries: symbiotic --
             stars: individual (RS Ophiuchi) --
             novae, cataclysmic variables --
             X-rays: binaries 
               }

   \maketitle
%
\section{Introduction}
\label{intro}

RS Ophiuchi (RS Oph) is a symbiotic recurrent nova that went into its
latest outburst on 2006 February 12.83 UT (\citealt{2006IAUC.8671....2N});
previous outbursts were recorded in 1898, 1933, 1958, 1967, and
1985\footnote{Two possible outbursts were also observed in 1907
(\citealt{2004IAUC.8396....2S}) and 1945 (\citealt{1993JAVSO..22..105O}).}
(see \citealt{1987rorn.conf....1R, 1987rorn.conf...27R} and references
therein). RS Oph is thought to be a binary system, comprising a
red giant star that does not fill its Roche lobe, and a white dwarf
of mass near the Chandrasekhar limit (\citealt{1994AJ....108.2259D,
1996ApJ...456..717S, 2000AJ....119.1375F}); the system has a period of
$455.72\pm 0.83$ days (\citealt{2000AJ....119.1375F}). Current theory
tells us that, in this class of binaries, the outbursts occur on the
white dwarf due to thermonuclear runaway of hydrogen-rich material
transferred from the companion red giant onto the surface of the
white dwarf (e.g. \citealt{1991ApJ...369..471K, 1999A&A...344..177A,
2000NewAR..44...81S}).

The strong interest in studying RS Oph originates mainly from the belief
that recurrent novae are progenitors of Type Ia supernovae (SNe Ia). Two
possible scenarios have been proposed: i) matter from the red giant
is accreted by the white dwarf, causing the latter to increase in mass
until it reaches the Chandrasekhar limit and explodes as an SN Ia; ii)
two white dwarf stars merge, with the combined mass momentarily exceeding
the Chandrasekhar limit, causing an SN Ia explosion. Which of these two
scenarios is the more plausible is still under debate, so studying the
nova outbursts from RS Oph can provide important insight into SNe Ia.

During the 2006 outburst an intensive international observing campaign
was organized, incorporating observations ranging from radio to X-ray
wavelengths, and monitoring the outburst since the early phases of its
evolution. The results included the detection of X-ray emission from
hot gas, evolving from a dominant plasma temperature of $\approx 10$~KeV
few days after the eruption (\citealt{2006Natur.442..276S}) to $\approx
4$~KeV 10 days after optical maximum (\citealt{2006ApJ...652..629B,
2006Natur.442..276S}). It has been suggested that bright X-ray
emission arises from the shock-heated extended outer atmosphere of
the red giant (e.g. \citealt{1985MNRAS.217..205B, 2006ApJ...652..629B,
2008ApJ...673.1067N, drake07}).

\chandra/HETG observations at day 13.9 revealed a rich spectrum of
emission lines indicative of emitting plasma with temperatures ranging
between $3$ and $60$~MK (\citealt{2008ApJ...673.1067N, drake07}).
\citet{drake07} noted that the lines are too strongly peaked to be
explained by a spherically-symmetric shock, suggesting a collimation
mechanism of the X-ray emitting plasma in the direction perpendicular
to the line-of-sight. The lines also appear asymmetric and slightly
blue-shifted, while the red wings of the line profiles become weaker
with increasing wavelength. \citet{drake07} suggested that the asymmetric
nature of the circumstellar medium (CSM) in which the explosion occurred
can be responsible for both the broad range in plasma temperature and
the shock collimation observed.

Here we investigate the origin of the line asymmetries, broadening and
blue-shifts observed with \chandra/HETG. We aim at exploring possible
diagnostics of the early blast wave and of the inhomogeneous CSM in
which the explosion occurred. To this end, we model the expansion of
the blast wave from the 2006 outburst of the recurrent nova RS Oph
through the extended outer atmosphere of the companion red giant,
using detailed 3-D hydrodynamic simulations. From the simulations we
synthesize the X-ray emission and derive the spectra as they would be
observed with \chandra/HETG.

In Sect. \ref{sec2} we describe the hydrodynamic model, the numerical
setup, and the synthesis of X-ray emission; in Sect. \ref{sec3} we discuss
the results; and finally in Sect. \ref{sec4} we draw our conclusions.

\section{Model}
\label{sec2}

\subsection{Hydrodynamic modeling}
\label{sec2.1}

The blast wave is modeled by numerically solving the time-dependent fluid
equations of mass, momentum, and energy conservation in a 3-D Cartesian
coordinate system $(x,y,z)$, taking into account the radiative losses
from an optically thin plasma and thermal conduction (including the
effects of heat flux saturation):

\begin{equation}
\begin{array}{l}\displaystyle
\frac{\partial \rho}{\partial t} + \nabla \cdot \rho \mbox{\bf u} = 0~,
\\ \\ \displaystyle
\frac{\partial \rho \mbox{\bf u}}{\partial t} + \nabla \cdot \rho
\mbox{\bf uu} + \nabla P = 0~,
\\ \\ \displaystyle
\frac{\partial \rho E}{\partial t} +\nabla\cdot (\rho E+P)\mbox{\bf u}
= -\nabla\cdot q -n_{\rm e} n_{\rm H} \Lambda(T)~.
\end{array}
\label{mod_eq}
\end{equation}

\[
\mbox{Here \hspace{0.5cm}} E = \epsilon +\frac{1}{2} |\mbox{\bf u}|^2~,
\]

\noindent
is the total gas energy (internal energy, $\epsilon$, and kinetic energy),
$t$ is the time, $\rho = \mu m_H n_{\rm H}$ is the mass density, $\mu
= 1.3$ is the mean atomic mass (assuming cosmic abundances), $m_H$
is the mass of the hydrogen atom, $n_{\rm H}$ is the hydrogen number
density, $n_{\rm e}$ is the electron number density, {\bf u} is the
gas velocity, $T$ is the temperature, $q$ is the conductive flux, and
$\Lambda(T)$ represents the radiative losses per unit emission measure
(e.g. \citealt{rs77}; \citealt{mgv85}; \citealt{2000adnx.conf..161K}). We
use the ideal gas law, $P=(\gamma-1) \rho \epsilon$, where $\gamma=5/3$
is the adiabatic index.

To allow for a smooth transition between the classical and saturated
conduction regime, we followed \citet{1993ApJ...404..625D} and defined
the conductive flux as (see also \citealt{2005A&A...444..505O})

\begin{equation}
q = \left(\frac{1}{q_{\rm spi}}+\frac{1}{q_{\rm sat}}\right)^{-1}~.
\end{equation}

\noindent
Here $q_{\rm spi}$ represents the classical conductive flux
(\citealt{spi62})

\begin{equation}
q_{\rm spi} = -\delta\rs{T}\epsilon 20\left(\frac{2}{\pi}\right)^{3/2}
\frac{(k\rs{B} T)^{5/2} k\rs{b}}{m\rs{e}^{1/2} e^4 Z \ln(\Lambda)}~\nabla T
\label{spit_eq}
\end{equation}

\noindent
where $k\rs{B}$ is the Boltzmann constant, $m\rs{e}$ is the electron
mass, $e$ is the electron charge, $Z$ is the average atomic number,
$\ln(\Lambda)$ is the Coulomb logarithm, $\delta\rs{T}$ and $\epsilon$
parameters depend on the chemical composition, and in a proton-electron
plasma: $\delta\rs{T}= 0.225$ and $\epsilon = 0.419$. The saturated flux,
$q_{\rm sat}$, is (\citealt{cm77})

\begin{equation}
q_{\rm sat} = -\mbox{sign}\left(\nabla T\right)~ 5\phi \rho c_{\rm
s}^3,
\label{therm}
\end{equation}

\noindent
where $c_{\rm s}$ is the isothermal sound speed, and $\phi$ is a
number on the order of unity. We set $\phi = 0.3$ according to the
values suggested for a fully ionized cosmic gas: $0.24<\phi<0.35$
(\citealt{1984ApJ...277..605G}; \citealt{1989ApJ...336..979B},
\citealt{2002A&A...392..735F}, and references therein).

In order to trace the motion of the material ejected by the violent
eruption, we consider a passive tracer associated with the ejecta. To
this end, we add the equation

\begin{equation}
\frac{\partial C_{\rm cl}}{\partial t} + \nabla \cdot C_{\rm cl} \mbox{\bf
u} = 0
\label{tracer}
\end{equation}

\noindent
to the standard set of hydrodynamic equations. $C_{\rm cl}$ is the
mass fraction of the ejecta inside the computational cell. The ejecta
material is initialized with $C_{\rm cl} = 1$, while $C_{\rm cl} = 0$
in the ambient medium. During the shock evolution, the ejecta and the
ambient medium mix together, leading to regions with $0 < C_{\rm cl} <
1$. At any time $t$ the density of ejecta material in a fluid cell is
given by $\rho_{\rm cl} = \rho C_{\rm cl}$.

The calculations described in this paper were performed using \FLASH,
an adaptive mesh refinement multiphysics code (\citealt{for00}). The
hydrodynamic equations are solved using the \FLASH\ implementation
of the piecewice-parabolic method (\citealt{cole84}). The code was
designed to make efficient use of massively parallel computers using the
message-passing interface (MPI) for interprocessor communications. The
code has been extended with additional computational modules
to handle radiative losses and thermal conduction (see
\citealt{2005A&A...444..505O} for the details of the implementation).

As an initial condition, we assume a spherical Sedov-type blast wave
with radius $r\rs{b0} = 1/3$~AU and with total energy $E\rs{b0}$,
originating from the thermonuclear explosion on the white dwarf. The
initial total energy of the explosion is partitioned so that 1/4
of the energy is contained in thermal energy, and the other 3/4 in
kinetic energy. The initial total mass of the ejecta is $M\rs{ej}$.
The blast propagates through the extended outer atmosphere (the wind)
of the companion red giant and is off-set from the origin of the wind
density distribution by 1.5 AU (i.e. the system orbital separation;
\citealt{1994AJ....108.2259D}). We follow \cite{2006Natur.442..279O} and
assume the gas density in the red giant wind proportional to $r^{-2}$
(where $r$ is the radial distance from the giant) and its temperature
$2\times 10^4$ K. Note that the temperature values of the red giant
wind (in symbiotic stars), $T\rs{w}$, can range between few $10^3$ K up
to $10^5$ K closer to the red giant (\citealt{1983A&A...126..407F}). To
evaluate the effects of $T\rs{w}$ on the results of our simulations,
we compared simulations with $T\rs{w} = 4000$ K with simulations with
$T\rs{w} = 20000$ K and found that the results do not change.

In addition to the $r^{-2}$ density distribution, we include
also an equatorial density enhancement (hereafter EDE) in
the red giant wind, as suggested by VLBA radio synchrotron
observations (\citealt{2006Natur.442..279O}) and by HST observations
(\citealt{2007ApJ...665L..63B}) of the 2006 blast wave, and as predicted
by detailed hydrodynamic modeling (\citealt{1999ApJ...523..357M,
2008A&A...484L...9W}). The mass density distribution of the unperturbed
CSM is given by:

\begin{equation} 
\rho = \frac{\rho\rs{w}}{r\rs{au}^2}+\rho\rs{eq}~
e^{-(h/L)^2}~, 
\label{csm}
\end{equation}

\noindent
where $\rho\rs{w} = \mu m_H n\rs{w}$ is the mass density at a distance
of 1 AU from the red giant, $r\rs{au}$ is the radial distance from the
giant in AU, $\rho\rs{eq} = \mu m_H n\rs{eq}$ is the density enhancement
at the equatorial plane, $h$ is the height above the equatorial plane
and $L$ is a characteristic length scale.

\begin{figure}[!t]
  \centering \includegraphics[width=7cm]{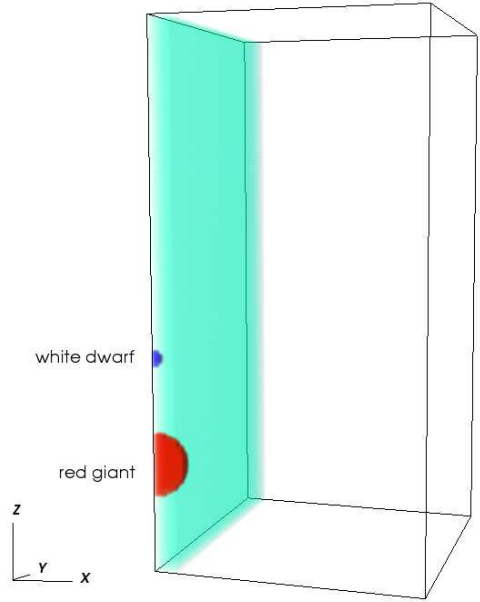}
  \caption{Sketch of the initial geometry of the RS Oph system. The
     figure shows the quadrant of the spatial domain that is modeled
     numerically. The red giant is at the origin of the coordinate system,
     $(x,y,z) = (0,0,0)$, and the computational domain extends 60 AU in
     the $z$ direction, and 30 AU in both the $x$ and $y$ directions;
     the white dwarf is located on the $z$ axis ($x=y=0$) at $z=1.5$
     AU. The $(y,z)$ plane is the equatorial plane and the green structure
     marks the equatorial density enhancement.}
  \label{fig1}
\end{figure}

In our 3-D simulations, the symmetry of the problem allows us to solve
the hydrodynamic equations in one quadrant of the spatial domain (see
Fig.~\ref{fig1}). The coordinate system is oriented in such a way
that both the white dwarf and the red giant lie on the $z$ axis. The red
giant is at the origin of the coordinate system, $(x,y,z) = (0,0,0)$,
and the computational domain extends 60 AU in the $z$ direction, and 30
AU in both the $x$ and $y$ directions; the white dwarf is arbitrarily
located to the north on the $z$ axis ($x=y=0$) at $z=1.5$ AU (i.e. the
system orbital separation; \citealt{1994AJ....108.2259D}).

At the coarsest resolution, the adaptive mesh algorithm used in the
\FLASH\ code (\PARAMESH; \citealt{mom00}) uniformly covers the 3-D
computational domain with a mesh of $4\times 4\times 8$ blocks, each
with $8^3$ cells. We allow for 8 nested levels of refinement during the
first 3 days of evolution and for 6 levels in the rest of the simulation,
with resolution increasing twice at each refinement level. The refinement
criterion adopted (\citealt{loehner}) follows the changes in density,
temperature, and tracer of ejecta. This grid configuration yields an
effective resolution of $\approx 0.03$ AU at the finest level during the
first 3 days of evolution (when the radius of the nova remnant was $< 5$
AU) and $\approx 0.1$ AU at later times, corresponding to an equivalent
uniform mesh of $1024\times 1024\times 2048$ and $256\times 256\times
512$ grid points, respectively. The higher spatial resolution during the
first 3 days of evolution is required to describe properly the development
of initially small structures dominated by radiative cooling. We use
reflecting boundary conditions at $x=x\rs{min}$ and $y=y\rs{min}$
(consistent with the adopted symmetry) and zero-gradient conditions at
the other boundaries.

We follow the expansion of the blast wave through the red giant
wind for $\approx 15$ days, considering two sets of simulations:
with or without the EDE. In both cases, we explore a parameter
space defined by the initial mass of ejecta $M\rs{ej}$ in the range
$10^{-7}-10^{-6}~M_{\odot}$, the initial energy of the explosion
$E_0$ in the range $10^{43}-10^{44}$~erg (namely the published
estimates for the ejected mass and energy of the explosion, e.g.
\citealt{2005ApJ...623..398Y, 2006Natur.442..276S, 2006Natur.442..279O}),
and the particle number density of the red giant wind at 1 AU
$n\rs{w}$ in the range $2\times 10^{7}-2\times 10^{10}$~cm$^{-3}$ (e.g.
\citealt{1994AJ....108.2259D}). In the set of simulations with the EDE,
we set $n\rs{eq}=10^{8}$~cm$^{-3}$ (consistent with the values found
by \citealt{2008A&A...484L...9W}) and we explore the case of $L=1$
and 2 AU (see Eq. \ref{csm}). Table \ref{tab1} summarizes the physical
parameters characterizing the simulations.

\begin{table}
\caption{Initial parameters of the 3-D simulations}
\label{tab1}
\begin{center}
\begin{tabular}{lccccc}
\hline
\hline
run  &  $M\rs{ej}^a$    & $E\rs{b0}^b$ & $n\rs{w}^c$ & $n\rs{eq}^d$ & $L^e$ \\
     &  [$M_{\odot}$] & [erg] & [cm$^{-3}$]  & [cm$^{-3}$] &  [AU] \\
\hline
ND-E43-N7  & $10^{-7}$ & $10^{43}$ & $2\times 10^{7}$ & 0 & - \\
ND-E44-N7  & $10^{-6}$ & $10^{44}$ & $2\times 10^{7}$ & 0 & - \\
ND-E43-N8  & $10^{-7}$ & $10^{43}$ & $2\times 10^{8}$ & 0 & - \\
ND-E44-N8  & $10^{-6}$ & $10^{44}$ & $2\times 10^{8}$ & 0 & - \\
ND-E43-N10 & $10^{-7}$ & $10^{43}$ & $2\times 10^{10}$ & 0 & - \\
ND-E44-N10 & $10^{-6}$ & $10^{44}$ & $2\times 10^{10}$ & 0 & - \\
YD-E43-N7-L2  & $10^{-7}$ & $10^{43}$ & $2\times 10^{7}$ & $10^{8}$ & 2 \\
YD-E44-N7-L2  & $10^{-6}$ & $10^{44}$ & $2\times 10^{7}$ & $10^{8}$ & 2 \\
YD-E43-N8-L2  & $10^{-7}$ & $10^{43}$ & $2\times 10^{8}$ & $10^{8}$ & 2 \\
YD-E44-N8-L2  & $10^{-6}$ & $10^{44}$ & $2\times 10^{8}$ & $10^{8}$ & 2 \\
YD-E44-N7-L1  & $10^{-6}$ & $10^{44}$ & $2\times 10^{7}$ & $10^{8}$ & 1 \\
\hline
\hline
\end{tabular}
\end{center}
$^a$ Initial mass of ejecta;\\
$^b$ Initial energy of the explosion;\\
$^c$ Particle number density of the red giant wind at 1 AU;\\
$^d$ Particle number density enhancement at the equatorial plane;\\
$^e$ Characteristic length scale in Eq. \ref{csm}.
\end{table}

\subsection{Synthesis of the X-ray emission}
\label{synthesis}

From the model results, we synthesize the X-ray emission originating from
the blast wave, applying a methodology analogous to the one described by
\cite{orlando2} in the context of the study of supernova remnants (see
also \citealt{2006A&A...458..213M}). The results of numerical simulations
are the evolution of temperature, density, and velocity of the plasma in
one quadrant of the whole spatial domain. We reconstruct the 3-D spatial
distribution of these physical quantities in the whole spatial domain,
and we rotate the system about the $z$ axis in such a way that the
inclination of the binary orbit to the line-of-sight (LoS) is 35$^{\rm
o}$, according to the values estimated from radial velocity observations
before the 2006 outburst (\citealt{1994AJ....108.2259D}). Both the red
giant and the white dwarf lie on the $z$ axis, and such a configuration
assumes that the 2006 explosion occurred at quadrature, in agreement
with the system ephemeris of \citet{2000AJ....119.1375F} (see also
\citealt{2006Natur.442..279O}).

The emission measure in the $j$-th domain cell is $\em\rs{j} = n\rs{Hj}^2
V\rs{j}$ (where $n\rs{Hj}^2$ is the hydrogen number density in the cell,
$V\rs{j}$ is the cell volume, and we assume fully ionized plasma). From
the values of emission measure and temperature in the cell, we synthesize
the corresponding X-ray spectrum, using the Astrophysical Plasma Emission
Code (APEC; \citealt{2001ApJ...556L..91S}) of hot collisionally ionized
plasma. We assume solar metal abundances of \cite{1998SSRv...85..161G}
(hereafter GS) for the CSM and abundances enhanced by $\times 10$
for the ejecta material (as suggested by \chandra/HETG observations;
\citealt{drake07}). The spectral synthesis takes into account the thermal
broadening of emission lines and the Doppler shift of lines due to the
component of plasma velocity along the LoS.

The X-ray spectrum from each cell is filtered through the absorption
column density relative to the position of the cell in the spatial
domain. The absorption consists of two components: 1) the photoelectric
absorption by the ISM, assuming a column density $N_{\rm H} =
2.4\times10^{21}$ cm$^{-2}$ (according to the value determined from
HI~21~cm measurements, e.g. by \citealt{1986ApJ...305L..71H}, and
consistent with a distance to RS Oph of $D\rs{oph} = 1.6$ kpc, e.g.
\citealt{1987rorn.conf..241B}) and 2) the local absorption by the
shocked CSM (with GS abundances) and by the ejecta (with GS abundances
$\times 10$) encountered within the blast wave. The ISM, CSM, and ejecta
absorption components are computed using the absorption cross-sections
as a function of wavelength from \cite{1992ApJ...400..699B}.

We integrate the absorbed X-ray spectra from the cells in the whole
spatial domain. The resulting X-ray spectrum (originating from
the whole blast wave) is then folded through the \chandra/HETG
instrument response, considering the source at a distance $D\rs{oph}
= 1.6$ kpc. The exposure time is assumed to be $t_{\rm exp} = 10$ ks
(the same value of the \chandra/HETG observations of RS Oph
analyzed by \citealt{2008ApJ...673.1067N, drake07}). 

\section{Results}
\label{sec3}

\subsection{Hydrodynamic evolution}

\begin{figure*}[!t]
  \centering \includegraphics[width=15.cm]{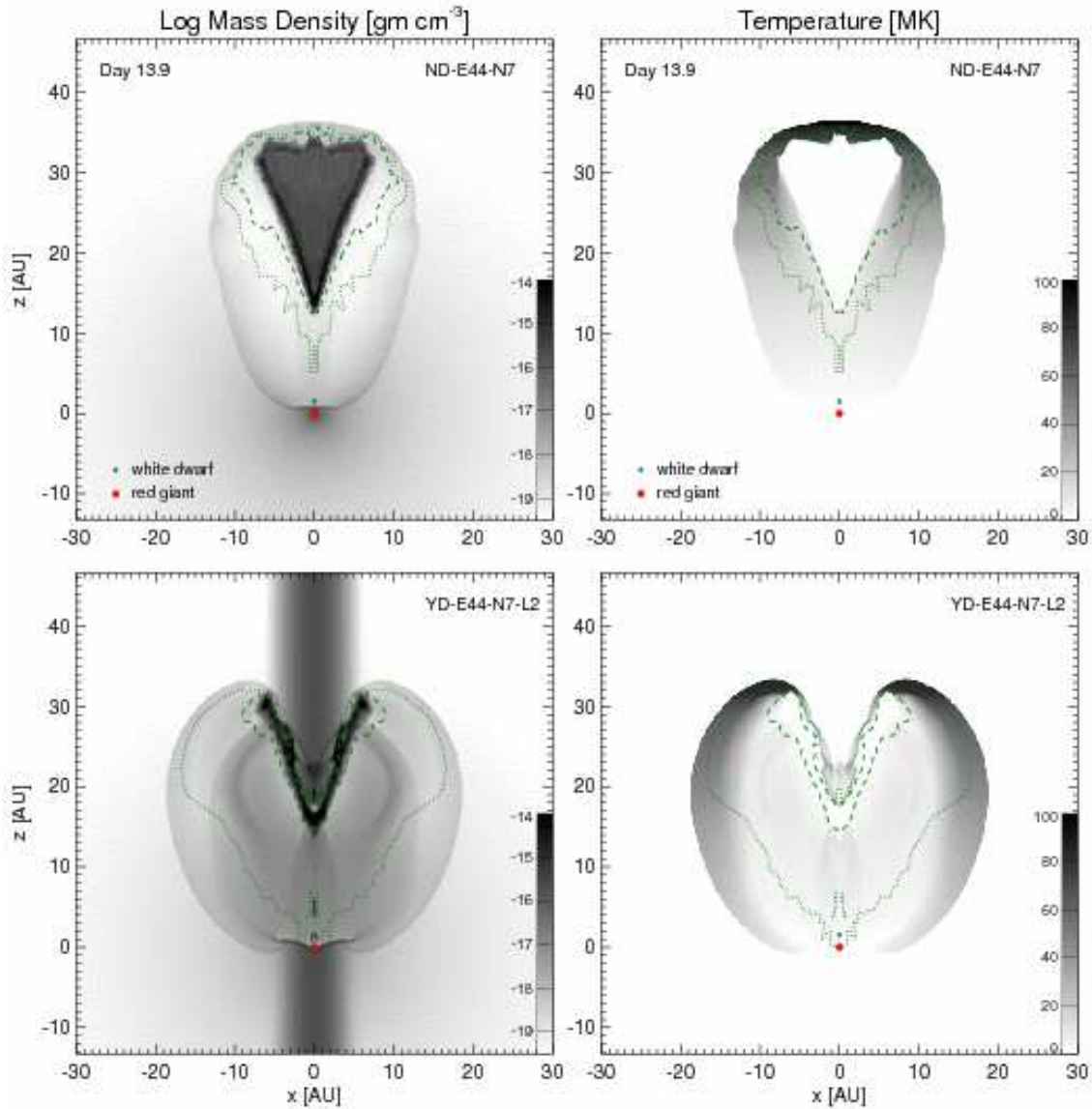}
  \caption{2-D sections in the $(x,z)$ plane of the mass density
  distribution, in log scale, (on the left) and of the temperature (on
  the right) in the simulations ND-E44-N7 (upper panels) and YD-E44-N7-L2
  (lower panels) at day 13.9. The EDE lies in the $(y,z)$ plane. The
  contours enclose zones whose content is made up of original ejecta
  material by more than 10\% (dotted) and 90\% (dashed).}
\label{fig2}
\end{figure*}

In all the models examined, we found the typical evolution of radiative
shocks propagating through an inhomogeneous medium: the fast expansion
of the shock front with temperatures of several millions degrees and the
development of dense and cold regions dominated by radiative cooling,
as the forward and reverse shocks progress through the CSM and ejecta,
respectively.  As examples, Fig.~\ref{fig2} shows 2-D sections in the
$(x,z)$ plane of the distributions of mass density (on the left) and of
temperature (on the right) for the models ND-E44-N7 (i.e. without the EDE)
and YD-E44-N7-L2 (with the EDE) at day 13.9.

The radiative cooling is very effective in the inner portion of the
remnant (dark regions in left panels of Fig.~\ref{fig2}) made up
mostly of ejecta material. Hot post-shock regions are dominated by
thermal conduction that contrasts the radiative cooling and suppresses
hydrodynamic instabilities that would develop during the evolution
of the blast wave (see \citealt{2005A&A...444..505O}). In particular,
the thermal conduction suppresses the Rayleigh-Taylor instabilities that
would develop at the contact discontinuity and that would drive the mixing
of ejecta material with the shocked CSM. In our simulations, the mass
mixing mainly comes from ejecta evaporation driven by thermal conduction
rather than from hydrodynamic ablation (see \citealt{2005A&A...444..505O,
2008ApJ...678..274O}). Fig.~\ref{fig2} also shows the distribution of ejecta
within the blast wave: regions whose content is made up of original
ejecta material by more than 90\% (dashed contours in the figure) are
mostly dense, cold and dominated by radiative cooling, whereas regions
made up of a mixture of ejecta and CSM are hot and dominated by thermal
conduction (see right panels of Fig.~\ref{fig2}). In all the cases, we
found aspherical shock morphologies, with the blast wave propagating
through inhomogeneous circumstellar gas distribution and being partially
refracted around the red giant star.

In models without EDE, the aspherical morphology is due to the fact
that the origin of the blast wave is off-set from the origin of
the wind density distribution by 1.5 AU. This off-set is very effective
in collimating the blast wave and ejecta. The shock front initially
propagating away (toward) from the red giant goes through the $r^{-2}$
($r^2$) density gradient and, as a consequence, its propagation velocity
and temperature are high (small): the blast wave appears elongated in
the direction away from the red giant and the hottest plasma component
emitting in the X-ray band lies to the side away from the red giant (see
upper panels of Fig.~\ref{fig2}). The shock front propagating toward
the red giant is strongly reflected by the high density medium and the
resulting reflected shock contributes to sweep out the ejecta material
away from the red giant (and from the white dwarf).

In models including the EDE, the aspherical shock morphology is quite
complex and originates from the propagation of the shock through both the
off-set red giant wind and the density enhancement at the equatorial plane
(see also \citealt{1987A&A...183..247G, 2008A&A...484L...9W}). The latter
component determines the collimation of the blast wave perpendicularly
to the plane of the orbit of the central binary system and leads to
a bipolar shock morphology distorted (by the off-set red giant wind)
and converging on the side away from the red giant (see lower panels
of Fig.~\ref{fig2}). In this case, the hottest X-ray emitting plasma
component lies in the two poles to the side away from the red giant (see
lower right panel of Fig.~\ref{fig2}). A movie showing the evolution
of 3D spatial distribution of mass density, in log scale, for the model
YD-E44-N7-L2 is provided as on-line material: the point of view is along
the $y$ axis during the whole evolution, than it first rotates about the
$x$ axis and finally about the $z$ axis up to an inclination of the binary
orbit of 35$^{\rm o}$ (i.e. the value estimated before the 2006 outburst).

\subsection{Emission measure vs. temperature}
\label{emt}

\begin{figure*}[!t]
  \sidecaption
  \includegraphics[width=12cm]{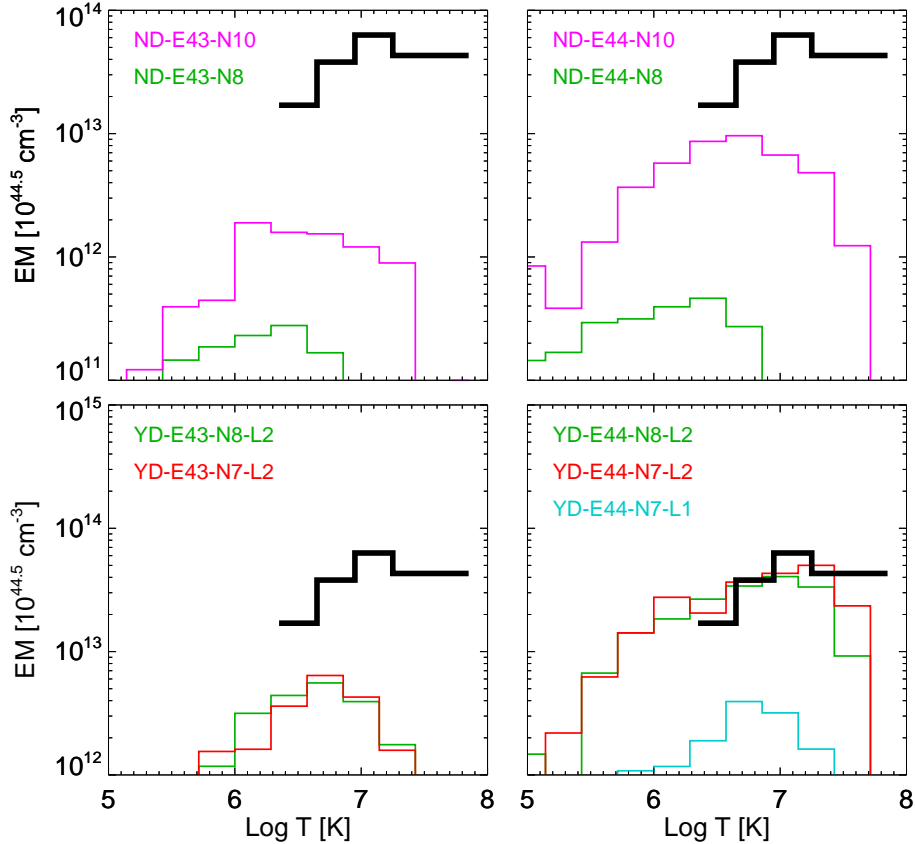}
  \caption{Emission measure vs. temperature distribution, EM($T$), of the
     blast wave at day 13.9. Upper panels show the result for models
     without EDE, lower panels for models with EDE. The black
     histogram represents the average EM($T$) estimated from the
     analysis of \chandra/HETG observations (\citealt{drake07}). The
     factor $10^{44.5}$ assumes a distance for RS Oph of 1.6 kpc.}
  \label{fig3}
\end{figure*}

From the model results, we derived the EM($T$) distribution of the blast
wave. As discussed in Sect. \ref{synthesis}, from the distribution of
mass density, we derive the emission measure in the $j$-th domain cell
as $\em_{\rm j} = n_{\rm Hj}^2 V_{\rm j}$. The EM($T$) distribution
is then derived by binning the emission measure values into slots of
temperature; the range of temperature [$4 < \log T (\mbox{K}) < 8$] is
divided into 15 bins, all equal on a logarithmic scale. Fig.~\ref{fig3}
shows the EM($T$) for models without EDE (upper panels), and with
EDE (lower panels) at day 13.9, in the range of temperature [$5 <
\log T (\mbox{K}) < 8$], and assuming a distance for RS Oph of 1.6 kpc
(\citealt{2007ApJ...665L..63B}). The figure also shows the average EM($T$)
estimated from resonance lines of abundant He- and H-like ions observed
with \chandra/HETG (\citealt{drake07}).

In models without EDE, the shape of the EM($T$) is characterized
by a bump centered at temperatures between 1 and 5 MK (see upper
panels in Fig.~\ref{fig3}). Increasing the initial energy of the
explosion $E\rs{b0}$ from $10^{43}$ to $10^{44}$ erg or increasing
the density of the red giant wind at 1~AU from $2\times 10^7$ to
$2\times 10^{10}$~cm$^{-3}$, increases the EM($T$) distribution at all
temperatures. However, in all the cases examined, the values of emission
measure at temperatures around 10~MK are orders of magnitude lower than
those derived from the observations (\citealt{drake07}) and most of
the shocked plasma is at temperatures below 5~MK. An initial energy of
the explosion $E\rs{b0} > 10^{45}$ erg (and an ejected mass $M\rs{ej}
> 10^{-5}~M_{\odot}$) and/or a density of the red giant wind at 1 AU
$n\rs{w} > 2\times 10^{10}$~cm$^{-3}$ (corresponding to a mass loss
rate of $> 5\times 10^{-6}~M_{\odot}$~yr$^{-1}$ for a wind velocity of
30 km s$^{-1}$) would probably reproduce the observed emission measure,
leading to a bump around 10~MK in the EM($T$) distribution. However,
these values of $E\rs{b0}$ and $M\rs{ej}$ are somewhat high for the true
energy and ejected mass of the explosion (\citealt{2005ApJ...623..398Y})
and the value of $n\rs{w}$ is also unrealistically large for the true
density of the red giant wind. We conclude that models without the
EDE fail to reproduce the values of emission measure derived from
observations.

In models including the EDE, the emission measure at temperatures above
5~MK is larger than that derived from the corresponding models without
EDE. The shocked plasma of the EDE leads to a bump around 10~MK in the
EM($T$) distribution (see lower panels in Fig.~\ref{fig3}).  The emission
measure of the bump depends on the initial energy of the explosion
$E\rs{b0}$. In models with $E\rs{b0}= 10^{43}$ erg, the EM around 10~MK
is much lower than that derived from observations. Indeed the observed
EM values can be reproduced with the models with $E\rs{b0}= 10^{44}$
erg. The shape of the EM($T$) distribution depends very slightly on the
density of the red giant wind: decreasing $n\rs{w}$, the EM slightly
increases at temperatures above $\approx 10$~MK. On the other hand,
the EM($T$) distribution strongly depends on the thickness of the EDE:
the model with the thinnest EDE predicts EM values at 10~MK well below
those observed. We found that model YD-E44-N7-L2 is our "best-fit"
model for reproducing the average EM($T$) estimated from observations
(see lower right panel in Fig.~\ref{fig3}).

Note that our models underestimate the emission measure values observed
at temperatures $> 20$~MK. The ``observed'' EM($T$) distribution at high
temperatures is mainly derived from the analysis of Fe\,XXV ($\lambda
1.85$), assuming GS abundances. However, \cite{drake07} noted that EM
values derived from Fe ions appear to be anomalous and suggested that Fe
probably suffers from effects of non-equilibrium ionization. In addition,
as discussed in the next section, we found that a large contribution to
the emission in the Fe\,XXV ($\lambda 1.85$) line comes from the shocked
ejecta material, with GS abundances $\times 10$. As a consequence,
the Fe~XXV EM derived by \cite{drake07} would be overestimated.

\subsection{X-ray emission}

\begin{figure}[!t]
  \centering \includegraphics[width=8.5cm]{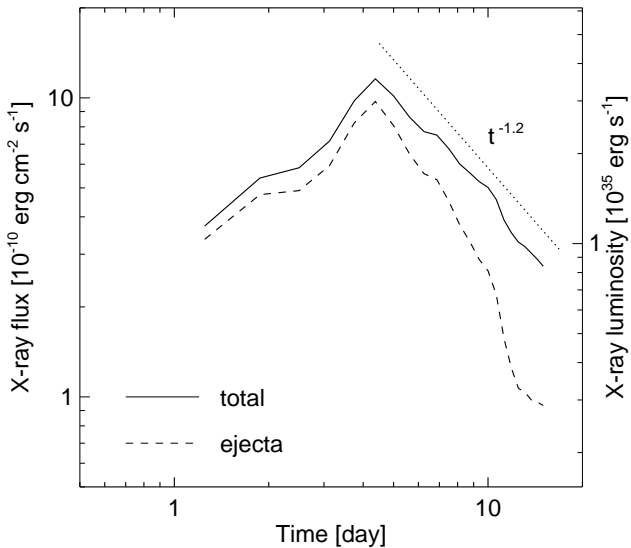}
  \caption{X-ray light curves in the $[0.6-12.4]$ keV band originating
           from the nova outburst (solid line) and from ejecta heated by
           the reverse shock (dashed line) for YD-E44-N7-L2. The X-ray
           flux is obtained by integrating the synthetic X-ray emission
           in the whole spatial domain and takes into account the effects
           of local absorption by shocked CSM and ejecta. The reference
           dotted line is a power law with index $\alpha = -1.2$.
           The X-ray luminosity is reported on the right axis assuming
           a distance to RS Oph of 1.6 kpc.}
\label{fig0}
\end{figure}

From the simulations, we synthesized the X-ray emission in the
$[0.6-12.4]$ keV band to match the observing range of \chandra/HETG,
using the method outlined in Sect.~\ref{synthesis}. Fig.~\ref{fig0}
shows the X-ray light curve in this band for YD-E44-N7-L2 (i.e. our
"best-fit" model), obtained by integrating the synthetic X-ray emission
in the whole spatial domain and taking into account the effects of
local absorption by shocked CSM and ejecta. The figure shows the X-ray
flux, $F_{\rm X}$, originating from the nova outburst together with the
contribution to X-rays from ejecta heated by the reverse shock (dashed
line). The X-ray flux reaches its maximum quite early, around day 4,
and then decays as $t^{-1.2}$. The time of maximum X-ray emission is
consistent with \swift/XRT observations (\citealt{2006ApJ...652..629B}),
although the observed X-ray flux is slightly higher\footnote{However,
note that the X-ray flux reported by \cite{2006ApJ...652..629B}
is corrected for the effects of absorption by intervening material,
whereas the synthesized X-ray flux takes into account the absorption
by shocked CSM and ejecta.} (by a factor $\approx 2$) than that
derived from the simulation and decays as $t^{-1.5}$ (see also
\citealt{2006Natur.442..276S}) instead of $t^{-1.2}$ as derived from the
simulation. The X-ray decay rate is mainly driven by radiative cooling,
which has a significant impact on the post-shock temperature structure
(\citealt{1987MNRAS.228..277O}). The contribution to X-ray luminosity
from shocked ejecta is rather important during the early phases of the
evolution (more than 80\% for $t< 5$ days) and then decreases down to 34\%
at day 15 due to significant radiative cooling.

\begin{figure}[!t]
  \centering 
  \includegraphics[width=7.5cm]{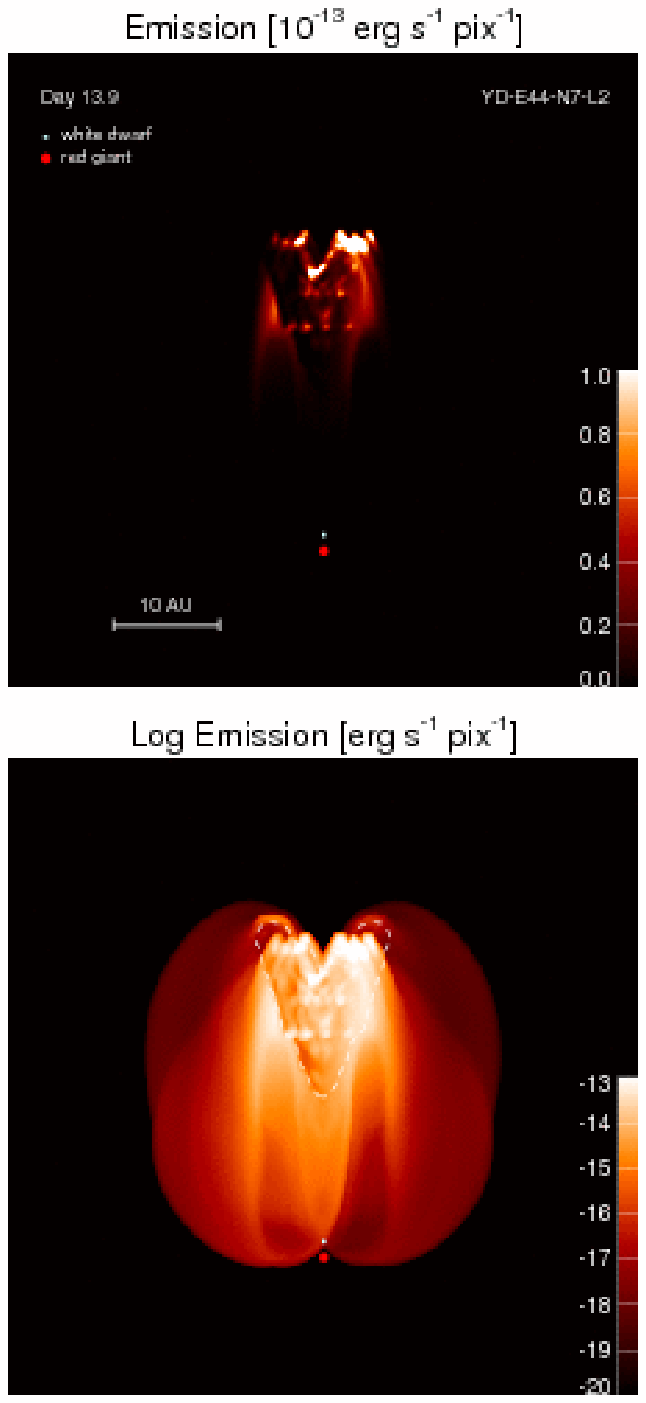}
  \caption{X-ray images in the [$0.6-12.4$] keV band in linear (upper
      panel) and logarithmic (lower panel) scales derived from model
      YD-E44-N7-L2 at day 13.9. The plane of the orbit of the central
      binary system lies on the $(y,z)$ plane and is assumed
      to be inclined by 35$^{\rm o}$ to the line-of-sight
      (\citealt{1994AJ....108.2259D}).  The dashed contour encloses the
      ejecta material leading to the
      largest photoelectric absorption.}
  \label{fig4}
\end{figure}

Fig.~\ref{fig4} shows the map of X-ray emission integrated along the
LoS and on pixels with size $\sim0.1$~AU at day 13.9 for YD-E44-N7-L2.
The map is drawn in linear (upper panel) and logarithmic (lower panel)
scales to highlight structures with very different emission levels. The
plane of the orbit of the central binary system lies on the $(y,z)$
plane and is assumed to be inclined by 35$^{\rm o}$ to the LoS,
consistent with the inclination of the orbit from orbital solutions
(\citealt{1994AJ....108.2259D}; see also Sect.~\ref{synthesis}). A movie
showing the evolution of X-ray emission for YD-E44-N7-L2 is also provided
as on-line material: the dotted gray line in the movie marks the position
of the forward shock.

\begin{figure*}[!t]
  \centering \includegraphics[width=16.cm]{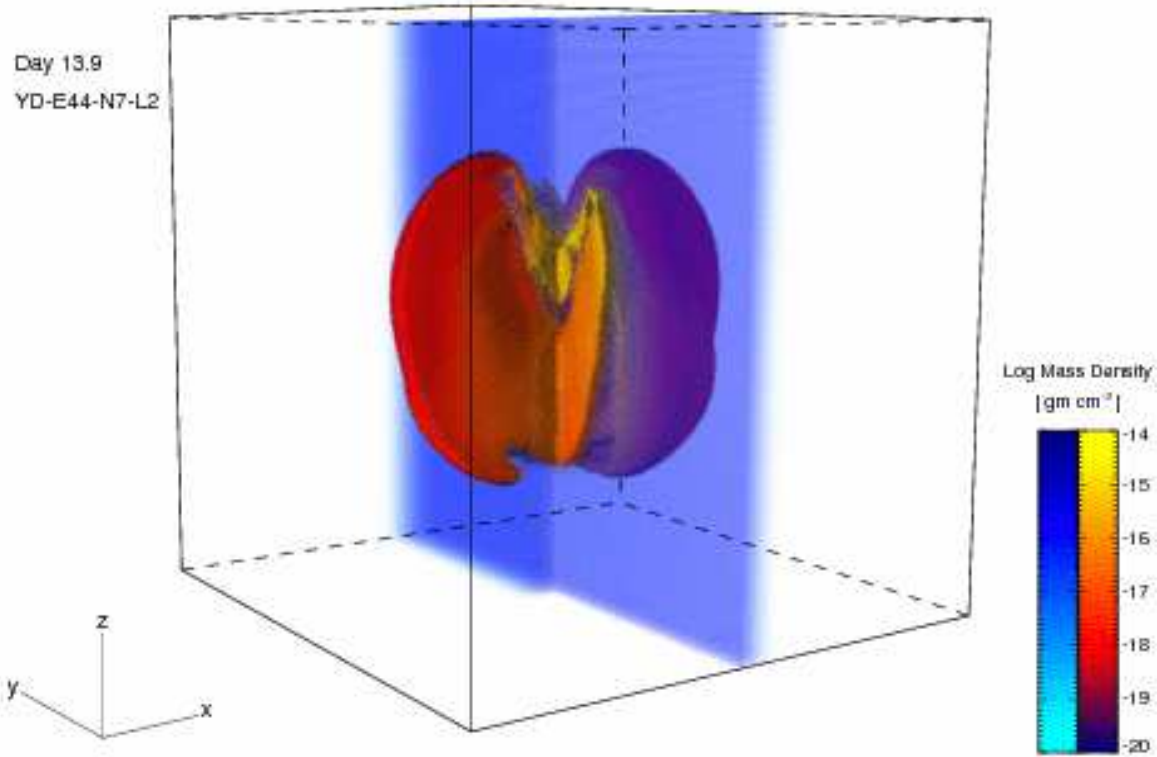}
  \caption{Three-dimensional rendering of mass density, in log scale,
     at day 13.9 for the model YD-E44-N7-L2. One quarter of the volume has
     been removed to show the internal structure of the blast wave. Red
     (blue) color table marks the plasma with temperature larger (lower)
     than 1~MK. The plane of the orbit of the central binary system
     lies on the $(y,z)$ plane.}
  \label{fig5}
\end{figure*}

Fig.~\ref{fig5} shows a 3-D rendering of mass density, in log scale,
at day 13.9 for YD-E44-N7-L2. We use two different color tables to
identify the plasma with temperature larger (red color table) or lower
(blue) than 1~MK. The volume has been clipped to show the internal
structure of the blast wave. The comparison between Fig.~\ref{fig4}
and Fig.~\ref{fig5} allows us to link X-ray emitting structures with
plasma structures originating from the interaction of the blast wave
with the inhomogeneous CSM.

Most of the X-ray emission originates from an irregular jet-like
structure, lying on the $(y,z)$ plane, with a size of $\approx 10$~AU
(see Fig.~\ref{fig4}). The structure is located to the side away from
the red giant, and the X-ray emission is maximum in a region with size
$\approx 4$~AU. By comparing the X-ray images in Fig.~\ref{fig4} with
the 3-D rendering of mass density in Fig.~\ref{fig5}, we note that
the X-ray emitting structure corresponds to a region just behind the
interaction front between the blast wave and the EDE (yellow region in
Fig.~\ref{fig5}); the post-shock plasma at the equator is characterized
by high density values ($10^{-15}< \rho < 10^{-13}$~gm~cm$^{-3}$) and
temperatures ranging between 1 and 50~MK, with the highest values of
density and temperature to the side away from the red giant.

The lower panel in Fig.~\ref{fig4} shows that faint X-ray emission also
arises from a double-lobed structure with a size of $\approx 30$~AU,
surrounding the X-ray emitting jet. By comparing Fig.~\ref{fig4} and
Fig.~\ref{fig5}, we found that the double-lobed structure corresponds to
regions downstream of the forward shock propagating perpendicularly to
the plane of the binary orbit. These regions are characterized by density
values ranging between $10^{-18}$~gm~cm$^{-3}$ and $10^{-17}$~gm~cm$^{-3}$
and temperatures between 10 and 85~MK. If the orientation of the
EDE is the same as discussed by \cite{2006Natur.442..279O}, we
speculate that this double-lobed structure roughly corresponds to the
synchrotron-emitting partial shell observed in high-resolution radio
observations (\citealt{2006Natur.442..279O, 2008ApJ...688..559R})
although X-ray and radio brightness distributions may be not
co-spatial\footnote{Note that synchrotron radio brightness distributions
also depend strongly on gradients of ambient magnetic field strength or
of ambient plasma density (\citealt{2007A&A...470..927O}).}.

\subsubsection{Line profile analysis}

We integrated the emission of the whole spatial domain and derived the
spectra for the HEG and MEG traces as predicted to be observed with the
\chandra/HETG, in order to compare our analysis with the results obtained
by \citet{2008ApJ...673.1067N} and by \citet{drake07}. As expected, the
synthetic spectra show emission lines from different elements, forming
over a wide range of plasma temperatures, and reflecting the broad nature
of the plasma temperature distribution discussed in Sect. \ref{emt}.

\begin{figure}[!t]
  \centering \includegraphics[width=8.cm]{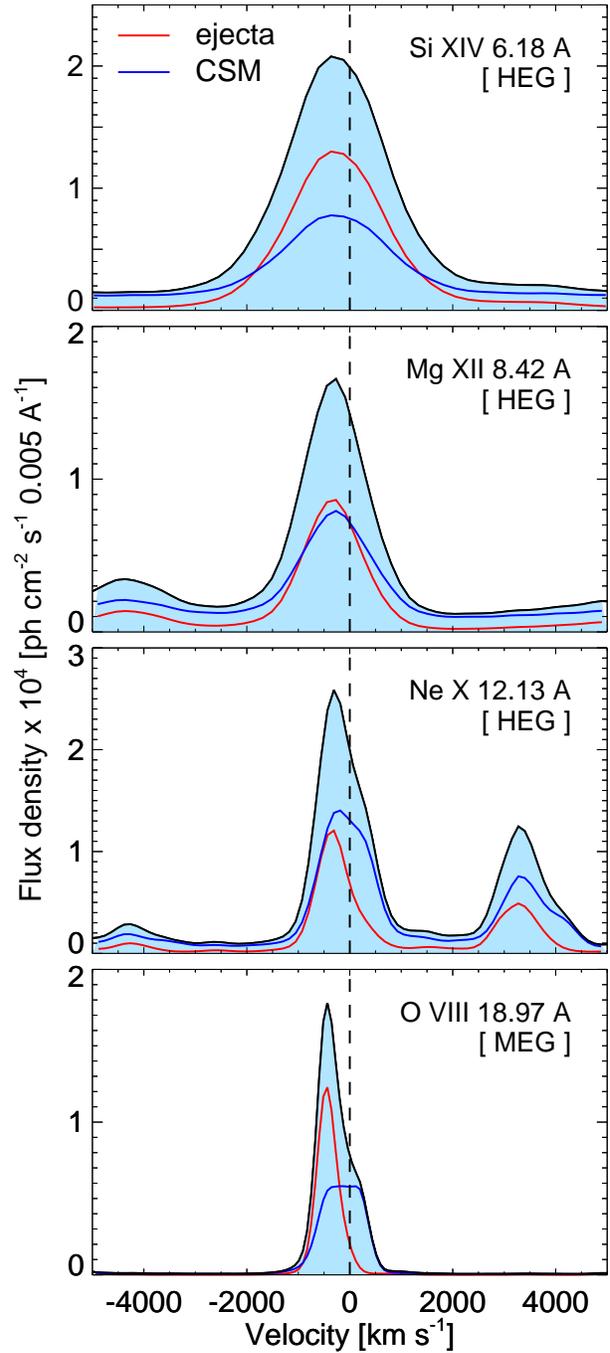}
  \caption{Synthetic velocity profiles for the H-like resonance lines
      of Si XIV, Mg XII, Ne X, and O VIII derived from model YD-E44-N7-L2
      at day 13.9. Si, Mg, and Ne lines are observed by the HEG, O
      line by the MEG. The figure also shows the contribution of shocked
      ejecta material (red lines) and shocked CSM (blue lines) to the
      X-ray emission.}
  \label{fig6}
\end{figure}

Following \cite{drake07}, we analyzed the line profiles of the most
prominent spectral lines to investigate the origin of the broadening
and asymmetries revealed in the \chandra/HETG observations (see,
also, \citealt{2008ApJ...673.1067N}). Analogously to \cite{drake07},
we restrict our analysis to HEG profiles, except in the case of O\,VIII
($\lambda 18.97$) which falls outside of the HEG range and is observed
by the MEG. Table~\ref{tab2} summarizes the results of our analysis for
the abundant He and H-like ions, together with the Fe\,XVII ($\lambda
15.01$) resonance line. Fig.~\ref{fig6} shows the line profiles for
the abundant H-like ions Si\,XIV ($\lambda 6.18$), Mg\,XII ($\lambda
8.42$), and Ne\,X ($\lambda 12.13$) observed by the HEG and O\,VIII
($\lambda 18.97$) observed by the MEG. Note that the line profiles are
affected by the instrument profile that is known to be more effective
for decreasing wavelengths.

The line profiles exhibit broadening and asymmetries remarkably similar
to those observed with \chandra/HETG. In velocity terms, most of the
synthetic lines exhibit FWHM ranging between 1000 and 3000 km s$^{-1}$
and FWZI ranging between 2500 and 8000 km s$^{-1}$, in agreement with
the results of \citet{2008ApJ...673.1067N} and \citet{drake07} (and
similar to those observed in the IR by \citealt{2006ApJ...653L.141D,
2007MNRAS.374L...1E}). The only exception is the bright He-like Fe\,XXV
($\lambda 1.85$) resonance line which exhibits much larger full-widths due
to substantial contribution to the emission from the other two components
of the Fe\,XXV triplet at $\lambda\approx 1.86$~\AA\ and $\lambda\approx
1.87$~\AA and due to strong instrument effects.

\citet{drake07} noted that the observed line profiles are more peaked
than expected for a spherically-symmetric shock and suggested that
the emission is strongly enhanced in the direction of the plane of the
sky. Our model predicts that most of the X-ray emission originates in
a compact region propagating away from the red giant in the direction
perpendicular to the LoS (see Fig.~\ref{fig4}). As a result, we found
that the line profiles are peaked as those observed.

In most of the lines, the centroids are systematically blue-shifted and
the amount of the shift depends on the wavelength (see Fig.~\ref{fig6}
and Table~\ref{tab2}): from the Si\,XIII ($\lambda 6.64$) line to the
H-like O\,VIII ($\lambda 18.97$) doublet, the amount of blue-shift
increases up to its maximum value. A similar trend has been found
by \citet{2008ApJ...673.1067N} from the analysis of \chandra/HETG
and XMM-Newton/RGS spectra of RS Oph at day 13.9, although the amount
of blue-shift found by these authors is systematically higher than that
predicted by our model. At variance with the other lines, the Fe\,XXV
($\lambda 1.85$) resonance line exhibits a significant red-shift. Again,
this different behavior is due to contribution to the emission from the
other two components of the Fe\,XXV triplet.

As a result of the systematic blue-shift of emission lines, the
line profiles tend to be more extended to the blue than the red (see
Fig.~\ref{fig6} and the line profile blue-shift and red-shift at zero
intensity in Table~\ref{tab2}) with the only exception of the Fe\,XXV
($\lambda 1.85$) line. As in the \chandra/HETG spectra, BSZI and RSZI are
observed to decrease with increasing wavelength, and this is particularly
evident in the O\,VIII ($\lambda 18.97$) line profile in the MEG spectrum.

\begin{table}
\caption{Parameters (in units of [km s$^{-1}$]) characterizing the
profiles of the most prominent spectral lines in the HEG and MEG
spectra$^a$}
\label{tab2}
\begin{center}
\begin{tabular}{lrrrrr}
\hline
\hline
     &  $v\rs{ctr}^b$  &  FWHM$^c$  & FWZI$^d$  & BSZI$^e$  & RSZI$^f$ \\
\hline
Fe\,XXV\,($\lambda 1.85$)   & 1410 & 11351 & 29188 & 13183 & 16004 \\
S\,XVI\,($\lambda 4.72$)    & -190 &  3172 &  8882 &  4631 &  4250 \\
S\,XV\,($\lambda 5.03$)     &  -30 &  2380 &  6843 &  3302 &  3540 \\
Si\,XIV\,($\lambda 6.18$)   & -291 &  2183 &  5580 &  2959 &  2620 \\
Si\,XIII\,($\lambda 6.64$)  & -203 &  1578 &  4510 &  2458 &  2052 \\
Mg\,XII\,($\lambda 8.42$)   & -160 &  1246 &  3561 &  1941 &  1620 \\
Mg\,XI\,($\lambda 9.16$)    & -196 &   981 &  2942 &  1667 &  1275 \\
Ne\,X\,($\lambda 12.13$)    & -222 &   988 &  2471 &  1458 &  1013 \\
Fe\,XVII\,($\lambda 15.01$) & -309 &   499 &  1297 &  1008 &   289 \\
O\,VIII\,($\lambda 18.97$)  & -355 &   553 &  1343 &  1066 &   276 \\
\hline
\hline
\end{tabular}
\end{center}
$^a$ Line profiles include the instrument profile that is known to be
more effective for decreasing wavelengths;\\
$^b$ Shift of the line centroid (negative values are for blue-shift);\\
$^c$ Full widths at half maximum;\\
$^d$ Full widths at zero intensity;\\
$^e$ Line profile blue-shift at zero intensity;\\
$^f$ Line profile red-shift at zero intensity;
\end{table}

The dashed contour in the lower panel in Fig.~\ref{fig4}
encloses the ejecta material leading to the largest photoelectric
absorption\footnote{Eq.~\ref{tracer} allows us to trace the ejecta
material in the CSM.}. Given the geometry of the system and the
inclination of the orbit, this material mostly absorbs the emission
originating from the receding portion of the X-ray emitting jet-like
structure (see also upper panel in Fig.~\ref{fig4}) that contributes to
the red-shifted emission. In fact, the figure shows an area of attenuated
X-ray emission located in the left lobe of the bipolar structure. The
resulting absorption of this component explains the asymmetry and the
systematic blue-shift of line profiles predicted by our model.

We also used the tracer associated with the ejecta to determine the
contribution of shocked ejecta to the X-ray emission (see red lines in
Fig.~\ref{fig6}). We found that the red-shifted emission of the ejecta
component is, in general, more absorbed than that of shocked CSM:
the asymmetry of observed line profiles is mainly due to the ejecta
component. We also found that, in general, the smaller the wavelength
the larger is the contribution of shocked ejecta to the emission; in
particular, the shocked ejecta contribute $\approx 60$\% to the emission
of Fe\,XXV ($\lambda 1.85$) line and $\approx 40$\% to the emission of
Fe\,XVII ($\lambda 15.01$) line. The different contributions of ejecta
to X-ray emission (together with possible deviations from equilibrium
of ionization) could explain the anomalous EM values derived from Fe
ions by \cite{drake07}. Among the lines analyzed, that with the largest
contribution of shocked ejecta (more than 70\%) is the S\,XV ($\lambda
5.03$) line.

\section{Summary and conclusions}
\label{sec4}

We have investigated through detailed hydrodynamic modeling the origin
of asymmetries and broadening of the emission lines observed with
\chandra/HETG during the 2006 outburst of RS Oph. To our knowledge,
these simulations represent the first attempt to model the 3D structure
of the blast wave originating from a nova outburst that simultaneously
considers radiative cooling and thermal conduction (including heat
flux saturation). Our findings have significant implications on the
diagnostics of the blast wave during the early phases of evolution and
lead to several useful conclusions:

\begin{enumerate}
\item In all the cases, we found that the nova remnant is highly
aspherical, with the morphology of the blast wave deeply influenced by
the pre-existing inhomogeneous CSM. Even in models without the EDE, the
off-set of the nova explosion from the origin of the wind density
distribution causes an effective collimation of the blast wave and
ejecta.

\item Models without the EDE cannot reproduce the values of emission
measure derived from \chandra/HETG observations unless the outburst
energy is $E\rs{b0} \gg 10^{45}$~erg (and the ejected mass in the
explosion $M\rs{ej} \gg 10^{-5}~M_{\odot}$) and/or the density of the
red giant wind is $n\rs{w} \gg 10^{10}$~cm$^{-3}$ at 1 AU; however,
for the physical parameters characterizing RS Oph, these values
seem too high for the true outburst energy and ejected mass in the
explosion (\citealt{2005ApJ...623..398Y}) and for the density of the
red giant wind (\citealt{1994AJ....108.2259D}). Models including
the EDE reproduce in a natural way, without any further ad hoc
assumption, the values of emission measure derived from \chandra/HETG
observations if $E\rs{b0} \approx 10^{44}$~erg and $M\rs{ej} \approx
10^{-6}~M_{\odot}$.

\item Our ``best-fit'' model (YD-E44-N7-L2) predicts that, at day 13.9,
most of the X-ray emission originates in a region with size $\approx 4$~AU
localized at the interaction front between the blast wave and the EDE to
the side away from the red giant; such an X-ray emitting region propagates
in the direction perpendicular to the LoS. As a result, the synthetic
line profiles are more peaked than expected for a spherically-symmetric
shock in nice agreement with the observations.

\item The synthetic line profiles are asymmetric and slightly blue-shifted
and they are remarkably similar to those observed. We found that the
observed asymmetries are due to substantial X-ray absorption of red-shifted
emission by ejecta material, confirming the conclusion of \cite{drake07}.

\item Both shocked CSM and shocked ejecta contribute to the observed X-ray
emission. The asymmetry and blue-shift of emission lines are mainly due to
the shocked ejecta component which is more affected by X-ray absorption
of red-shifted emission. In general, the contribution of shocked ejecta
to the X-ray emission decreases with increasing wavelength.

\end{enumerate}

Our model shows that the broad range in the plasma temperature and the
asymmetries of line profiles observed in \chandra/HETG spectra are due
to the interaction of the blast wave with the pre-existing inhomogeneous
CSM. In addition, the asymmetric nature of the CSM into which the early
blast wave is driven is also responsible of the apparent shock collimation
in the plane of the sky. Note that the jet-like ejection emitting
in the X-ray band predicted by our model does not coincide with the
synchrotron jet observed in the radio band (\citealt{2006Natur.442..279O,
2008ApJ...688..559R}): the former propagates in the plane of the orbit of
the central binary system, whereas the latter propagates perpendicularly
to the same plane.

Our analysis confirms that the comparison of high quality data with
detailed hydrodynamic modeling can be a powerful tool to study and
diagnose the physical properties of outbursts from recurrent novae and to
interpret the observations. In particular, our results may provide useful
constraints on the circumstellar gas distribution of RS Oph and suggest
the existence of a dense region of the red giant wind at the equatorial
plane. We also show that most of the early X-ray emission arises from
this region as the blast wave from the 2006 outburst decelerates.

The scenario of an equatorial density enhancement is consistent with
observations of the 2006 outburst in other wavelength bands, for instance
VLBA radio synchrotron observations of an evolving ring-like structure
(\citealt{2006Natur.442..279O, 2008ApJ...688..559R}) and HST observations
of a bipolar nebular structure (\citealt{2007ApJ...665L..63B}). Also, the
existence of the equatorial density enhancement may explain the presence
of dust in the circumstellar environment in which the explosion occurs
(\citealt{2007MNRAS.374L...1E, 2008ApJ...677.1253B}).

Our findings are also consistent with the asymmetric blast wave
evolution described analytically by \cite{1987A&A...183..247G} and
with the results of hydrodynamic models describing the evolution of the
circumstellar environment in binary systems (\citealt{1999ApJ...523..357M,
2008A&A...484L...9W}). In particular, in systems comprising a red giant
star, these models predict an equatorial density enhancement created by
a spiral shock wave caused by the motion of the stars through the cool
wind of the red giant; the enhancement is most pronounced in systems
with the smallest binary separation, comparable to that in RS Oph.

It is worth noting that, in \cite{2008A&A...484L...9W}, the blast
wave appears more collimated perpendicularly to the plane of the orbit of
the central binary system than in our simulations. The reason is that the
CSM immediately around the white dwarf (within $1/3$~AU from the dwarf)
is characterized by a rather complex density structure (see the lower
right panel of Fig. 2 of \citealt{2008A&A...484L...9W}) determining the
strong collimation of the blast wave, at the very beginning of the nova
eruption (not described by our model). Since we assume the initial blast
wave to be spherical (see Sect. \ref{sec2.1}), our model does not take
into account such a strong initial collimation of the shock. To evaluate
the effect of the initial condition on the evolution of the blast wave,
we compared our ``best-fit" model with an additional simulation with an
identical setup but assuming an initial ellipsoid-shaped blast with the
major axis perpendicular to the orbital plane. Such an initial condition
describes the strong shock collimation occurred during the early phases
of the nova eruption. Taking into account this initial shock collimation,
we found that the shock expands very rapidly toward the perpendicular
to the orbital plane and, after few days of evolution, its shape is
similar to that found by \cite{2008A&A...484L...9W}. Nevertheless,
we also found that the main features characterizing the X-ray emission
(broadening, asymmetries, blue-shift of emission lines) arising from the
blast wave do not change qualitatively. Therefore, the main conclusions
of this paper on the origin of the X-ray emission and of asymmetries
and broadening of emission lines observed with \chandra/HETG remain.

The comparison of our model results with \chandra/HETG data suggests
that the mass of ejecta in the 2006 outburst was of the order of
$10^{-6}~M_{\odot}$. From the analysis of the spectra collected with
the Rossi X-ray Timing Explorer (RXTE), \cite{2006Natur.442..276S}
derived the ejecta mass to be of a few times $10^{-7}~M_{\odot}$ on
the basis of some assumptions (time of transition to the Sedov-Taylor
phase, shock speed during the ejecta-dominated phase, mass density
inside the binary). However, taking into account the uncertainties
in the determination of these values, the mass of ejecta derived by
\cite{2006Natur.442..276S} can range between few times $10^{-7}~M_{\odot}$
and few times $10^{-6}~M_{\odot}$, i.e. a range of values consistent
with that derived by our model.

Considering the mass of ejecta in the 2006 outburst of the
order of $10^{-6}~M_{\odot}$, it comes out that the white dwarf
is increasing in mass if its growth rate is larger than $5 \times
10^{-8}~M_{\odot}$~yr$^{-1}$. For lower values of the growth rate, the
nova would throw off more mass than the white dwarf may have accreted
in the intervening 22 years. Recently \citet{2007ApJ...659L.153H} have
estimated the white dwarf mass in RS Oph to be $1.35 \pm 0.01~M_{\odot}$
and its growth rate to be about $10^{-7}~M_{\odot}$~yr$^{-1}$ in average
(see also \citealt{2008arXiv0807.1251K}). According to our results,
therefore, we conclude that the white dwarf mass is effectively growing
up and RS Oph could be the progenitor of a SN Ia as the white dwarf
reaches the Chandrasekhar limit.

It will be interesting to expand the present study, including the ambient
magnetic field, to investigate the evolution of the blast wave in later
evolutionary phases and to make predictions on the synchrotron radio
emission. The detailed comparison of model results with observations
may lead to a major advance in the study of interactions between the
blast wave and the magnetized CSM in recurrent novae, and may provide
important insight into SNe Ia, of which recurrent novae are believed
to be progenitors.

\bigskip
\acknowledgements{
This work was supported in part by the Italian Ministry of
University and Research (MIUR) and by Istituto Nazionale di Astrofisica
(INAF). The software used in this work was in part developed by the
DOE-supported ASC / Alliance Center for Astrophysical Thermonuclear
Flashes at the University of Chicago, using modules for thermal conduction
and optically thin radiation built at the Osservatorio Astronomico di
Palermo. The simulations were executed on the Grid infrastructure of
the Consorzio COMETA. This work makes use of results produced by the
PI2S2 Project managed by the Consorzio COMETA, a project co-funded
by the Italian Ministry of University and Research (MIUR) within the
Piano Operativo Nazionale ``Ricerca Scientifica, Sviluppo Tecnologico,
Alta Formazione'' (PON 2000-2006). More information is available at
http://www.pi2s2.it and http://www.consorzio-cometa.it.}

\bibliographystyle{aa}
\bibliography{references}

\end{document}